\documentclass[%
 reprint,
superscriptaddress,
 amsmath,amssymb,
 aps,
 pre,
]{revtex4-1}
\usepackage{dsfont}
\usepackage[dvipsnames]{xcolor}
\usepackage{tikz}
\usepackage{comment}
\usetikzlibrary{backgrounds}
\usetikzlibrary{arrows,shapes}
\usetikzlibrary{tikzmark}
\usetikzlibrary{calc}
\usepackage{graphicx}% Include figure files
\usepackage{dcolumn}% Align table columns on decimal point
\usepackage{bm}% bold math
\usepackage[colorlinks= true, linkcolor=blue, citecolor=blue, urlcolor=blue]{hyperref}% add hypertext capabilities
\usepackage{lipsum}
\usepackage{xspace}
\usepackage{soul}

\begin{document}
%\preprint{APS/123-QED}

\title{Dynamic redundancy and mortality in stochastic search}  

%\date{\today}
\author{Samantha Linn}
\email{s.linn@imperial.ac.uk}
\affiliation{Department of Mathematics, Imperial College London, London SW7 2AZ, UK}

\author{Aanjaneya Kumar}
\email{aanjaneya@santafe.edu}
\affiliation{Santa Fe Institute, 1399 Hyde Park Road, Santa Fe, NM 87501, USA}
\affiliation{High Meadows Environmental Institute, Princeton University, Princeton, NJ, 08544, USA}

\begin{abstract}
Stochastic search processes are a fundamental part of natural and artificial systems. In such settings, the number of searchers is rarely constant: new agents may be recruited while others can abandon the search. Despite the ubiquity of these dynamics, their combined influence on search efficiency remains unexplored. Here we present a general framework for stochastic search in which independent agents progressively join and leave the process, a mechanism we term \emph{dynamic redundancy and mortality} (DRM). Under minimal assumptions on the underlying search dynamics, this framework yields exact first-passage time statistics. It further reveals surprising connections to stochastic resetting, including a regime in which the resetting mean first-passage time emerges as a universal lower bound for DRM, as well as regimes in which DRM search is faster. We illustrate our results through a detailed analysis of one-dimensional Brownian DRM search. Altogether, this work provides a rigorous foundation for studying first-passage processes with a fluctuating number of searchers, with direct relevance across physical, biological, and algorithmic systems.
\end{abstract}

\maketitle

\emph{Introduction.---}Stochastic search processes pervade natural and artificial systems -- from molecules locating binding sites and immune cells finding antigens, to animals foraging and algorithms exploring vast landscapes. In all such settings, performance depends on how efficiently a system locates a target under uncertainty. This efficiency is often quantified by \emph{first-passage times} (FPTs), defined as the time taken by a searcher to find a target for the first time \cite{redner2001guide}, which has received attention in applications ranging from chemical reaction kinetics to stochastic optimization \cite{weiss1967first,crandall1966some,szabo1980first,bray2013persistence,fauchald2003using,pulkkinen2013,metzler2014first,zhang2016first,iyer2016first,grebenkov2020preface}.

However, in all the above examples and many other real-world search processes, the population of searchers is not fixed -- new agents can be recruited to join the search (\emph{dynamic redundancy}) while existing ones drop out, die, or decay (\emph{mortality}). The acceleration of search through redundancy, or the presence of multiple searchers, has received wide attention \cite{SCHUSS201952,grebenkov2020single, lawley2020distribution, Sponsini_adv_2024, Hass2024, Maclaurin2025, targetsearch} and its dynamic counterpart, where searchers are stochastically injected into the system, has recently been introduced~\cite{campos2024dynamic,meyer2025optimal, tung2025first,tung2025passage,grebenkov2025fastest}. Moreover, search processes under mortality have a long history, where the effect of finite searcher lifetime has been shown to substantially impact FPTs \cite{yuste2013exploration, MeersonRedner2015,  meerson2015number, grebenkov2017escape, Ma2020, lawley2021fastinact, Radice2023, abad2024search}. These two ubiquitous features--dynamic redundancy and mortality--fundamentally reshape search statistics and rarely act in isolation, but their \emph{combined} influence on stochastic search remains unexplored. This key gap is the focus of this work.

\begin{figure}[h]
    \centering   \includegraphics[width=0.9\linewidth]{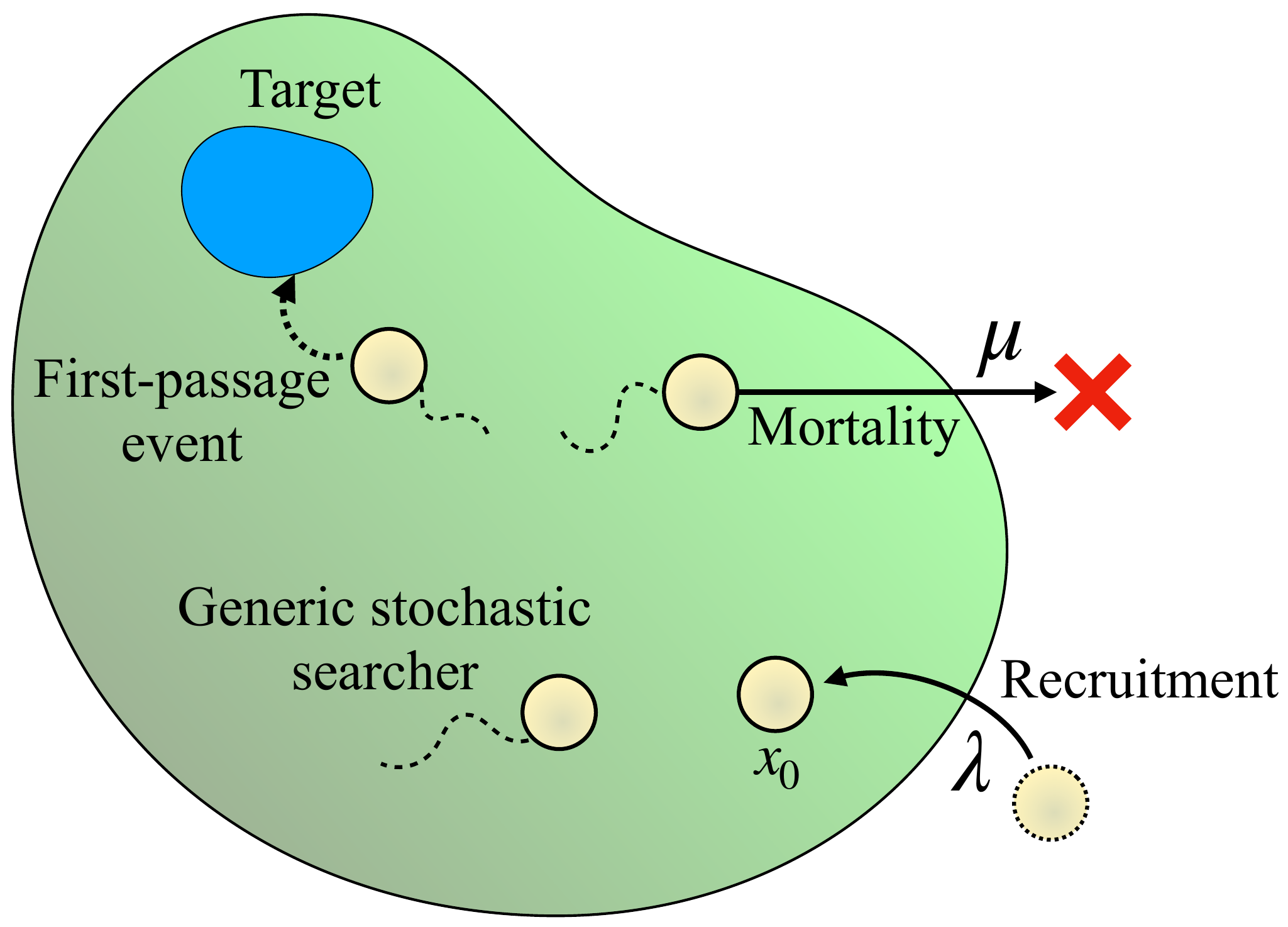}
    \caption{A schematic for the stochastic search process where searchers are recruited to the search at rate $\lambda$ (dynamic redundancy) and abandon the search process at rate $\mu$ (mortality). The central focus of this Letter is to characterize the time taken for the target to be found for the first time under dynamic redundancy and mortality (DRM).}
    \label{fig:schematic}
\end{figure}

In this \emph{Letter}, we characterize the combined impact of dynamic redundancy and mortality (hereafter referred to as DRM) on FPT statistics. Under minimal assumptions on the underlying search process, we explicitly express the DRM survival probability in terms of the survival statistics of a single mortal searcher. In doing so we reveal a subtle relationship between DRM and stochastic resetting. We find in particular that these processes, which are macroscopically identical, have nuanced differences on the scale of individual trajectories that manifest in the FPT statistics. The stochastic resetting mean FPT (MFPT) moreover serves as a universal lower bound for the DRM process with equal recruitment and mortality rates. We also establish an optimal upper bound for general recruitment and mortality rates, thereby inferring that the MFPT remains finite even when the average number of searchers at any given time is less than one, i.e.\ when mortality dominates over dynamic redundancy. We conclude with a detailed case study of a Brownian DRM search process in one-dimension (1D).

\emph{The setting.---}Consider a stochastic search process in an arbitrary domain $\Omega$ that is either confined or unbounded. One searcher, initially positioned at $x_0 \in \!\Omega$, seeks a target fixed at $\Omega^*\subset \Omega\backslash x_0$. At rate $\lambda$, new searchers are recruited to the search process, initialized at $x_0$, and independently search for the target. Each searcher independently drops out of the search at rate $\mu$. Figure \ref{fig:schematic} illustrates a schematic of this DRM search process. We note that this model was first introduced in Ref.~\cite{campos2024dynamic} by Campos and M\'endez who analyzed the case $\Omega= \mathbb{R}$ with the target at the origin and $\mu=0$.

The central quantity of interest in this \emph{Letter} is the time taken for the target to be found for the first time by any of the searchers, which we denote by $T_{\lambda,\mu}$. Note that when $\lambda\!=\!\mu\!=\!0$, then $T_{\lambda,\mu} \equiv T_{0,0} =: \tau$ is the classical FPT---the random time taken by a single searcher initially located at $x_0$ to find the target. Throughout this \emph{Letter}, we assume $\tau$ is strictly positive and not always infinite. These constitute our only assumptions on the underlying search process, which we emphasize are extremely mild and broadly applicable.

\emph{Failure of conventional FPT approach.---}A natural starting point for many first-passage problems is to study the evolution of the searcher density. As an illustration consider a Brownian DRM process in 1D. The mean density of searchers $\rho(x,t)$ evolves according to the Fokker-Planck equation
\begin{align} \label{pde}
    \frac{\partial \rho(x,t)}{\partial t} = D \frac{\partial^2 \rho(x,t)}{\partial x^2} - \mu \rho(x,t) + \lambda \delta(x-x_0)
\end{align}
where the first term on the right-hand side corresponds to the diffusive motion of the particles, and the second and third terms correspond to the loss and gain of density through mortality at $x$ and recruitment at $x_0$, respectively. To obtain the first-passage statistics, one conventionally introduces an absorbing boundary at the target and computes the flux through the target. In single searcher problems, the flux is exactly the FPT density. This is true, for instance, when $\lambda=\mu=r$ in Eq.\ \eqref{pde}, which describes a single Brownian particle undergoing stochastic resetting to $x_0$ at rate $r$. In DRM, however, the number of searchers dynamically fluctuates and so the FPT density does not equal, and cannot easily be obtained from, the flux through the target. 

\emph{An alternative approach to FPTs.---}The inability to extract the FPT density via standard boundary value methods, exemplified through the aforementioned example of 1D Brownian motion, necessitates an alternative approach which we develop here. We return to the more general DRM search setting. To obtain FPT statistics, consider the survival probability $S_{\lambda,\mu}(t):=\mathbb{P}(T_{\lambda,\mu}\! >\! t)$ defined as the probability that the target is not found by any searcher until time $t\!>\!0$. In Ref.~\cite{SI} we elaborate on methods introduced in Refs.~\cite{campos2024dynamic, tung2025first} to show that $S_{\lambda,\mu}(t)$ satisfies 
\begin{align} \label{SI}
    S_{\lambda,\mu}(t) = S_{0,\mu}(t) \exp\Big( \!-\!\lambda \int_0^t (1-S_{0,\mu}(t'))\,\textup{d}t'\Big)
\end{align}
where $S_{0,\mu}(t)$ denotes the probability that a single mortal searcher that dies at rate $\mu$ has not found the target by time $t$. We can express $S_{0,\mu}(t)$ as
\begin{align} \label{Smu}
    S_{0,\mu}(t) := \mathbb{P}(T_{0,\mu} > t) = 1 - \int_0^t e^{-\mu t'}\, \mathbb{P}(\tau=t')\, \textup{d}t'
\end{align}
where $\mathbb{P}(\tau=t)$ denotes the FPT density of a single searcher without mortality. While we focus on DRM as described by Eqs.\ \eqref{SI} and \eqref{Smu}, one could easily generalize the mortality distribution via substitution of the exponential function  or generalize the recruitment position by integrating over said position density within the integrand of Eq.\ \eqref{Smu}.

Equation \eqref{SI} provides an exact expression for the DRM survival probability of any search process whose underlying FPT satisfies the minimal assumptions on $\tau$. While Eq.~\eqref{SI} admits the full FPT statistics, we focus our analysis on the MFPT $\mathbb{E}[T_{\lambda,\mu}]$ which can be obtained as
\begin{equation}\label{mfpt}
    \mathbb{E}[T_{\lambda,\mu}] = \int_0^\infty S_{\lambda,\mu}(t) ~ \text{d}t.
\end{equation}
It is shown in Ref.~\cite{SI} that for large values of $\lambda$ and $\mu$ the coefficient of variation of the FPT is close to one, and for moderate to large values of $\lambda$ and $\mu$ the mean serves as a good descriptor of the FPT distribution.

\emph{Universal bounds on the DRM MFPT}.---We prove in Ref.~\cite{SI} that the MFPT satisfies the universal bounds
\begin{align} \label{bounds}
    \frac{1-p_\mu}{\lambda p_\mu} \leq \mathbb{E}[T_{\lambda,\mu}] \leq \frac{1-p_\mu}{\lambda p_\mu} - \frac{\partial }{\partial\mu} \log p_\mu
\end{align}
for all $\lambda,\mu>0$ where we define $p_\mu \in (0,1)$ by
\begin{align} \label{pk}
    p_\mu := \int_0^\infty e^{- \mu t}\, \mathbb{P}(\tau=t) \,\textup{d}t.
\end{align}
The quantity $p_\mu$ can be interpreted as the probability that an individual mortal searcher finds the target before dying. The bounds in Eq.~\eqref{bounds} admit a simple pathwise interpretation. Suppose the search process is not terminated when the target is first found, and index searchers in order of recruitment. The lower bound in Eq.~\eqref{bounds} is the mean recruitment time of the first successful searcher by recruitment index, while the upper bound is obtained by additionally waiting for this searcher’s hitting time, \emph{i.e.,} its MFPT conditioned on success \cite{SI}. In the true DRM process, since another searcher could find the target between this first-born successful searcher's recruitment and hitting times, the right-hand side is indeed an upper bound (and not an exact equality). In Ref.~\cite{SI}, we further demonstrate how the same bounds can be obtained from Eq.~\eqref{SI}. 

An important consequence of the upper bound is that the MFPT is finite for all $\lambda, \mu\!>\!0$. Even when $\lambda\!<\!\mu$, and hence there is less than one searcher on average at any given time, the MFPT remains finite due to the continuous recruitment of (new) searchers whose inter-recruitment times are almost surely finite. We further note that the upper bound expressed in Eq.~\eqref{bounds} is optimal -- it is attained for deterministic first-passage processes. If $\mathbb{P}(\tau=t)=\delta(t-t^*)$, then $\mathbb{E}[T_{\lambda,\mu}] =  \frac{1-p_\mu}{\lambda p_\mu} + t_*$ and thus the upper bound in Eq.~\eqref{bounds} holds with equality.

\emph{Stochastic resetting as a universal DRM lower bound when $\lambda \!=\! \mu$.---}Stochastic resetting is the process by which prescribed dynamics are randomly repositioned in the state space, often instantaneously and to the initial condition (hence `resetting') \cite{Evans2020,evans2011diffusion,evans2011diffusion2,pal2015diffusion,pal2016diffusion,reuveni_optimal_2016,bhat2016stochastic,pal2017first,chechkin2018random}. Consider a single stochastic searcher that undergoes such resetting at rate $r\!>\!0$. The corresponding FPT $\tau_r$ is known to satisfy \cite{FPTuFSR}
\begin{align} \label{ETr}
    \mathbb{E}[\tau_r] = \frac{1-p_r}{rp_r}.
\end{align}
\begin{figure}[t]
    \centering   \includegraphics[width=0.95\linewidth]{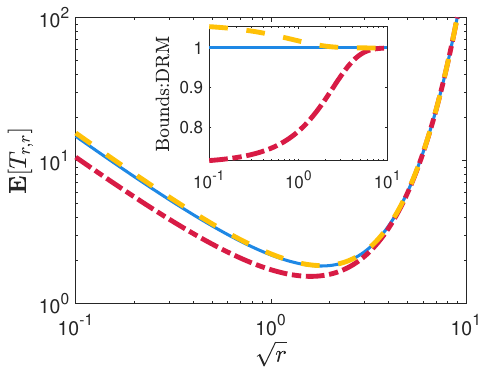}
    \caption{Balanced ($\lambda\!=\!\mu\!=\!r$) DRM MFPT $\mathbb{E}[T_{r,r}]$ for Brownian motion on the half line with $x_0\!=\!1$ and $D\!=\!1$. The solid blue curve indicates the solution computed from Eq.~\eqref{mfpt} via quadrature; the dash-dot red curve indicates the exact solution in resetting theory and the dashed yellow curve indicates the algebraically computed upper bound as stated in Eq.~\eqref{low}. The inset shows ratios of each bound to the MFPT curve.}
    \label{fig:bounds}
\end{figure}
Now consider when $\lambda\!=\!\mu\!=\!r$, which we call a \emph{balanced} DRM process. In this case, the lower bound to the DRM MFPT in Eq.~\eqref{bounds} is exactly the resetting MFPT in Eq.~\eqref{ETr}. This bound is rather surprising; when $\lambda=\mu$ there is an average of one searcher at any given time so one might expect the two processes to simply be equivalent. Moreover we have already seen through Eq.~\eqref{pde} the macroscopic equivalence between the DRM and resetting processes. Despite these similarities, resetting always outperforms balanced DRM. The discrepancy between their performances can however become arbitrarily small in the large $r$ limit. In particular, we show in Ref.~\cite{SI} that the second term in the upper bound of Eq.~\eqref{bounds} vanishes as $\mu=r\to\infty$ so long as a single searcher without mortality can reach the target arbitrarily quickly with positive probability (as is the case in Brownian motion). In this case, it follows that
\begin{align} \label{mean}
    \lim_{r\to\infty} rp_r\,\mathbb{E}[T_{r,r}] = \lim_{r\to\infty} rp_r\mathbb{E}[\tau_r] = 1.
\end{align}
This set of observations is one of the main results of this Letter.

\begin{figure*}[t]
    \centering   
    \includegraphics[width=\linewidth]{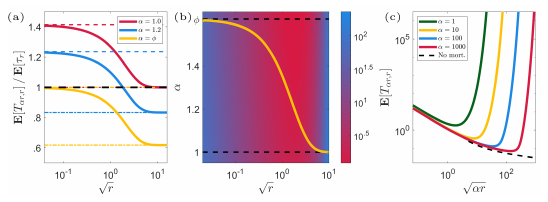}
    \caption{DRM for 1D Brownian motion with $\alpha\!>\! 0$ where $\mu\!=\! r$ and $\lambda\!=\!\alpha r$. (a) Ratio of DRM to resetting MFPTs denoted by solid curves computed numerically. The dash-dot lines are the high-turnover limits as in Eq.~\eqref{alpha_final}. The dotted lines are the low-turnover limits as in Eq.~\eqref{eq:1d_lowr_mfpt}. (b) Phase diagram and heatmap of the DRM MFPT. The solid yellow curve denotes the separatrix for the DRM and resetting MFPT; above (below) this curve the DRM MFPT is less (greater) than the stochastic resetting MFPT. The letter $\phi$ denotes the golden ratio. In Ref.~\cite{SI} we prove convergence of the small and large $r$ limits. (c) The no mortality result is an asymptotic result for the frequent recruitment limit recently derived in Ref.~\cite{tung2025first} and is extrapolated here for finite but large recruitment rates (dashed black line). Throughout, $x_0\!=\!1$ and $D\!=\!1$.}
    \label{fig:alphaone}
\end{figure*}

\emph{Case study of Brownian motion in 1D.---}We now illustrate our results through the example of Brownian motion in 1D. Consider a Brownian DRM search on the real line ($\Omega=\mathbb{R}$) with a searcher initially located at $x_0>0$ and a target placed at the origin ($\Omega^* =0$). Additional searchers are recruited to $x_0$ at rate $\lambda$ and each independently dies at rate $\mu$. We construct this DRM survival probability using Eq.~\eqref{SI} and the known density of $\tau$ \cite{redner2001guide},
\begin{align} \label{ft}
    \mathbb{P}(\tau=t) = \frac{x_0}{\sqrt{4D\pi t^3}} \,e^{\frac{-x_0^2}{4Dt}}
\end{align}
where $D>0$ is the diffusion coefficient. Substituting Eq.~\eqref{ft} into Eq.~\eqref{pk} yields $p_\mu = e^{-x_0\sqrt{\mu/D}}$, and the DRM survival probability $S_{\lambda,\mu}(t)$ is determined via further substitution into Eq.~\eqref{SI}. The full expression for $S_{\lambda,\mu}(t)$ is rather involved even for this simple case and is therefore contained in Ref.\ \cite{SI}. However, when this DRM process is balanced one can check that the large $r$ behavior of the survival probability exhibits the same exponential decay as that of stochastic resetting \cite{FPTuFSR,SI},
\begin{align} \label{lim}
    \lim_{r\to\infty} e^{rp_rt}\, S_{r,r}(t) = \lim_{r\to\infty} e^{rp_rt}\, \mathbb{P}(\tau_r > t) = 1.
\end{align}

One also finds through substitution the explicit bounds on the MFPT in this setting,
\begin{align} \label{low}
      \frac{ e^{x_0\sqrt{\mu/D}} \!-\! 1 }{\lambda} \leq \mathbb{E}[T_{\lambda,\mu}]  \leq \frac{ e^{x_0\sqrt{\mu/D}} \!-\! 1 }{\lambda} + \sqrt{\frac{x_0^2}{4D\mu}}.
\end{align}
Setting $\lambda\!=\!\mu\!=\!r$ yields
\begin{align} \label{low}
      \frac{ e^{x_0\sqrt{r/D}} \!-\! 1 }{r} \leq \mathbb{E}[T_{r,r}]  \leq \frac{ e^{x_0\sqrt{r/D}} \!-\! 1 }{r} + \sqrt{\frac{x_0^2}{4Dr}}.
\end{align}
We illustrate the results from Eqs.~\eqref{low} and \eqref{mean} in Fig.~\ref{fig:bounds}. Clearly, the lower bound diverges both when $r\!\to\! 0$ and $r\!\to\!\infty$. Moreover, since $\mathbb{E}[T_{r,r}]$ is finite for all finite $r\!>\!0$, we infer the existence of an `optimal' rate that minimizes the balanced DRM MFPT. 

For general, uncorrelated $\lambda,\mu\!>\!0$, reducing the MFPT merely involves maximizing $\lambda$ while minimizing $\mu$. Setting $\lambda\!=\!\mu\!=\!r$ introduces a trade-off between increasing redundancy and decreasing mortality and we see in Eq.~\eqref{low} that stochastic resetting always outperforms DRM. These observations raise a natural question: how much must redundancy dominate a DRM system such that it outperforms stochastic resetting?

\emph{When is DRM faster than resetting?}---Suppose $\mu\!=\! r$  and $\lambda \!=\!\alpha r$ where $\alpha\!> \!0$ and $r\!>\!0$ is the \emph{turnover} rate which sets a timescale for the DRM process. If $\alpha\!<\!1$, mortality dominates but the DRM MFPT remains finite. For $\alpha\!>\!1$ redundancy dominates and we find that mean DRM search times can be faster than those of stochastic resetting. In particular, we show in Ref.~\cite{SI} that for any $\alpha\!>\!1$ there exists a $r^*\!>\!0$ such that for all $r\!>\!r^*$ the DRM MFPT outpaces the corresponding resetting MFPT with resetting rate $r$. Furthermore, when the second term in the upper bound of Eq.~\eqref{bounds} vanishes,
\begin{align} \label{alpha_final}
    \lim_{r\to\infty} \frac{\mathbb{E}[T_{\alpha r,r}]}{ \mathbb{E}[\tau_r]} = \lim_{r\to\infty} rp_r\,\mathbb{E}[T_{\alpha r,r}] = 1/\alpha.
\end{align}
Equation~\eqref{alpha_final} asserts that for any $\alpha>1$ the DRM MFPT outpaces stochastic resetting by a factor $\alpha$ in the high-turnover limit.

Additionally, for the specific case of 1D Brownian motion, we show in Ref.~\cite{SI} that in the low-turnover limit, 
\begin{equation}\label{eq:1d_lowr_mfpt}
    \lim_{r\to 0} \frac{\mathbb{E}[T_{\alpha r,r}]}{ \mathbb{E}[\tau_r]} = \lim_{r\to 0} \sqrt{\frac{D \,r}{x_0^2}} \,\mathbb E[T_{\alpha r,r}] = \frac{\sqrt{\alpha + 1}}{\alpha}.
\end{equation} Consequently, for $\alpha > \phi:= (1+\sqrt{5})/2 \approx 1.618$,  1D Brownian DRM search outperforms stochastic resetting. We note that this critical value $\phi$ is exactly the golden ratio. Figure \ref{fig:alphaone}(a) demonstrates this set of results summarized by Eqs.~\eqref{alpha_final} and \eqref{eq:1d_lowr_mfpt}. Interestingly, we see that only moderately high turnover is required for minimally redundant-dominant (e.g.\ $\sim\!\alpha=1.2$) DRM systems to outperform stochastic resetting, while for $\alpha > \phi$, DRM outperforms resetting for all $r$. Figure \ref{fig:alphaone}(b) further illustrates the dependence of $\mathbb{E}[T_{\alpha r,r}]$ on $\alpha$ and $r$. Evident from the heatmap is that for any $\alpha$ there exists an optimal rate $r$ to minimize the MFPT. Moreover, the yellow curve acts as a separatrix in the phase space: for all pair values $(r,\alpha)$ that lie above it, DRM outpaces stochastic resetting whereas, for all values below, the resetting MFPT is a lower bound to the DRM MFPT.

\emph{Redundant-dominance differs from no-mortality.} Setting $\lambda/\mu\!=\!\alpha$ also allows us to establish the non-trivial role of mortality on the MFPT. Consider the large $\alpha$ limit (\emph{i.e.}, $\lambda\!\gg\!\mu$). One may initially expect the limiting results to resemble the case of frequent recruitment without mortality ($\mu=0$), which was recently studied under the nomenclature of `fast immigration' \cite{tung2025first}. In fact this behavior is not realized. Rather, as illustrated in Fig.~\ref{fig:alphaone}(c), the influence of mortality always eventually manifests, causing the MFPT to diverge. However, for modest $r$ and large $\alpha$, the MFPT trend mimics that of fast immigration.

\emph{Discussion.---}In this \emph{Letter}, we developed a probabilistic framework to study FPTs of stochastic search with dynamic redundancy and mortality. By expressing the full FPT survival probability in closed form, we obtained optimal model-independent upper and lower bounds on the MFPT under mild assumptions on the underlying search dynamics. A central result is that when recruitment and mortality occur at equal rates, stochastic resetting is a global lower bound for DRM. However, DRM can always outpace stochastic resetting even when recruitment only slightly dominates mortality for sufficiently high searcher turnover. We moreover found the redundant-dominant limit to exhibit remarkably different behavior than the corresponding no-mortality system. Altogether this work provides the first exact treatment of FPTs of stochastic search processes characterized by the continuous recruitment and mortality of searchers, with direct relevance across physical, biological, and algorithmic systems.

Our results open several directions for future work. An exciting next step would be to show that the FPT of \emph{any} balanced DRM search process satisfying our minimal assumptions does not just have mean $1/(r p_r)$, but is in fact a rate-$r p_r$ exponential random variable in the high-turnover limit. More broadly the framework developed here is readily extensible to other practically relevant first-passage problems, including stochastic gating \cite{scher2021unified, kumar_first_2021, scher2024continuous,kumar2023inference,toste2022arrival}, spatially distributed recruitment \cite{ro2017parallel,grebenkov2020single}, non-Markovian mortality processes \cite{stage2017aging}, and to higher-order statistics such as the $k$th fastest FPT \cite{weiss1983order,yuste2001order1,yuste2001order2,tung2025passage,lawley2020universal,grebenkov2022first}. Finally, while we focused on independent recruitment and mortality, it would be of interest to model them as correlated, where perfect correlation corresponds to stochastic resetting and non-correlation is akin to the work herein.

\emph{Acknowledgments.---}The authors are grateful to Paul Bressloff, Jos\'e Giral-Barajas, Sean Lawley, and Sid Redner for several helpful discussions, and to Yuval Scher for feedback on the manuscript. SL was supported by the U.S.\ National Science Foundation grant DMS-2503350. AK acknowledges the support of the Complexity Postdoctoral Fellowship of the Santa Fe Institute.

\clearpage
\onecolumngrid
\setcounter{page}{1}
\renewcommand{\thepage}{S\arabic{page}}
\setcounter{equation}{0}
\renewcommand{\theequation}{S\arabic{equation}}
\setcounter{figure}{0}
\renewcommand{\thefigure}{S\arabic{figure}}
\setcounter{section}{0}
\renewcommand{\thesection}{S\arabic{section}}
\setcounter{table}{0}
\renewcommand{\thetable}{S\arabic{table}}

\begin{center}
{\large \textbf{Supplemental Material for ``Dynamic redundancy and mortality in stochastic search"}}
\end{center}

\author{Samantha Linn}
\email{s.linn@imperial.ac.uk}
\affiliation{Department of Mathematics, Imperial College London, London SW7 2AZ, UK}
\author{Aanjaneya Kumar}
\email{aanjaneya@santafe.edu}
\affiliation{Santa Fe Institute, 1399 Hyde Park Road, Santa Fe, NM 87501, USA}
\affiliation{High Meadows Environmental Institute, Princeton University, Princeton, NJ, 08544, USA}

\maketitle

\onecolumngrid

This Supplemental Material provides further discussion and derivations that support the findings reported in the Letter.

% \tableofcontents

\section{Derivation of the DRM survival probability}
Here we outline the derivation of Eq.~(2) from the main text. To start, we express the DRM survival probability $S_{\lambda,\mu}(t)$ in terms of the survival probability of individual mortal searchers $S_{0,\mu}(t)$, which satisfies
\begin{align} \label{Smut}
     S_{0,\mu}(t) := \mathbb{P}(T_{0,\mu} > t) = 1 - \int_0^t e^{-\mu t'}\, \mathbb{P}(\tau=t')\, \textup{d}t'.
\end{align}
We note that Eq.~\eqref{Smut} accounts for the possibility of the searcher abandoning the search process before time $t\!>\!0$ as well as continuing to actively search for the target. To account for dynamic redundancy, we denote by $S_{\lambda,\mu}^{(n)}(t)$ the DRM survival probability conditioned on there being $n-1\!\geq\! 0$ recruitment events by time $t\!>\!0$ (not including the recruitment of the initial searcher at $t\!=\!0$). Since searchers are recruited by a (homogeneous) Poisson point process, the recruitment times of $n-1$ searchers are independently and uniformly distributed on $[0,t]$,
\begin{align} \label{Sn}
    S_{\lambda,\mu}^{(n)}(t) = S_{0,\mu}(t) \Big( \frac{1}{t} \int_0^t S_{0,\mu}(t')\,\textup{d}t' \Big)^{n-1}.
\end{align}
The exact number of recruitment events is Poisson distributed with mean $\lambda t$ and so the DRM survival probability $S_{\lambda,\mu}(t)$ can be understood in terms of $S_{\lambda,\mu}^{(n)}(t)$ via the law of total probability,
\begin{align} \label{Ssum}
    S_{\lambda,\mu}(t) = \sum_{n=1}^\infty \frac{e^{-\lambda t}(\lambda t)^{n-1}}{(n-1)!} S_{\lambda,\mu}^{(n)}(t).
\end{align}
Substitution of Eq.~\eqref{Sn} into Eq.~\eqref{Ssum} yields the final expression,
\begin{align} \label{final}
    S_{\lambda,\mu}(t) = S_{0,\mu}(t) \exp\left( \!-\lambda \int_0^t (1-S_{0,\mu}(t'))\,\textup{d}t'\right).
\end{align} 
which is Eq.~(2) of the main text. A similar derivation was carried out in Refs.~\cite{tung2025first} and \cite{campos2024dynamic}, for first-passage times under dynamic redundancy \emph{without mortality}.

\section{High-turnover behavior of the FPT coefficient of variation is unity}
\begin{figure}[t!]
    \centering   
    \includegraphics[width=.48\linewidth]{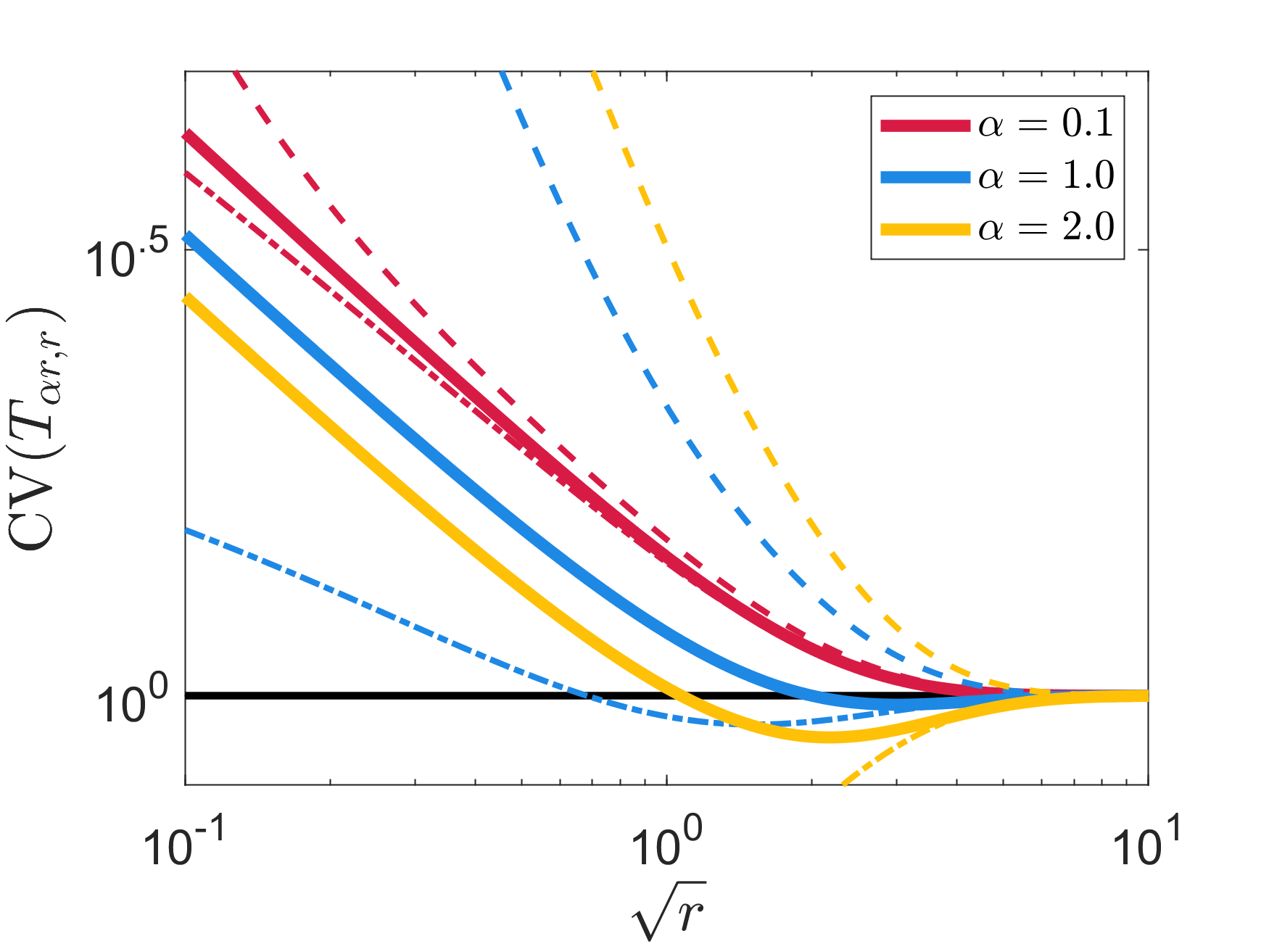}
    \caption{DRM MFPT coefficient of variation (CV) for Brownian motion on the half line with $\lambda=\alpha r$ and $\mu=r$. The solid curves indicate the CV computed from Eq.~\eqref{cv} via quadrature. The dash-dot curves indicate the lower bound as in Eq.~\eqref{frlow} and the dashed curves indicate the upper bound as in Eq.~\eqref{frupper}. Here, $x_0=1$ and $D=1$. Evidently, for moderate to large values of $r$, CV is close to $1$, and for large $r$, CV $\to 1$.}
    \label{fig:cv}
\end{figure}

Consider a DRM search process with $\lambda = \alpha r$ and $\mu=r$. We can describe the coefficient of variation (CV) of the DRM FPT $T_{\alpha r,r}$, which we denote by $CV(T_{\alpha r,r})$, in terms of the corresponding survival probability $S_{\alpha r,r}(t)$,
\begin{align} \label{cv}
    CV^2(T_{\alpha r,r}) + 1 = \frac{2\int_0^\infty tS_{\alpha r,r}(t)\,\textup{d}t}{\big( \int_0^\infty S_{\alpha r,r}(t)\,\textup{d}t \big)^2} =: f(r).
\end{align}
Below we establish that $f(r)\sim 2$ as $r\to\infty$ which implies that the CV converges to unity in the high-turnover limit. We have already shown that for all $r>0$
\begin{align} \label{avg}
    \frac{1-p_r}{\alpha rp_r} \leq \int_0^\infty S_{\alpha r,r}(t)\,\textup{d}t \leq \frac{1-p_r}{\alpha rp_r} - \frac{\partial}{\partial r} \log p_r.
\end{align}
Moreover, since $S_{\alpha r,r}(t) \geq (1-p_r)e^{-\alpha rp_rt}$ it follows that
\begin{align} \label{lowb}
    \int_0^\infty tS_{\alpha r,r}(t)\,\textup{d}t \geq \frac{1-p_r}{(\alpha rp_r)^2}.
\end{align}
Combining the upper bound in Eq.~\eqref{avg} with Eq.~\eqref{lowb}, we have
\begin{align} \label{frlow}
    f(r) \geq \frac{2(1-p_r)/(\alpha rp_r)^2}{\Big(\frac{1-p_r}{\alpha rp_r} - \frac{\partial}{\partial r} \log p_r\Big)^2}.
\end{align}
Taking $r\to\infty$ yields $\liminf_{r\to\infty} f(r) = 2$. The limit supremum is more involved. Recall from Eq.~\eqref{s0mu} that $S_{0,r}(t) = 1-p_r+R_r(t)$ with $R_r(t)$ as in Eq.~\eqref{Rmu}. Then
\begin{align} \label{intsum}
    \int_0^\infty tS_{\alpha r,r}(t)\,\textup{d}t = \frac{1-p_r}{p_r}\int_0^\infty t(p_r-R_r(t))G_r(t)\,\textup{d}t + \frac{1}{p_r}\int_0^\infty tR_r(t)G_r(t)\,\textup{d}t.
\end{align}
Using Eq.~\eqref{avg} and $G_r(t)$ as defined in Eq.~\eqref{eq:Gdef} we determine an upper bound on the first integral in Eq.~\eqref{intsum},
\begin{align}
    \int_0^\infty t(p_r-R_r(t))G_r(t)\,\textup{d}t = \frac{-1}{\alpha r} \int_0^\infty tG_r'(t)\,\textup{d}t
    = \frac{1}{\alpha r}\int_0^\infty G_r(t)\,\textup{d}t
    \leq \frac{1}{\alpha r}\Big( \frac{1}{\alpha r} + \frac{1-p_r}{\alpha rp_r} - \frac{\partial}{\partial r} \log p_r \Big).
\end{align}
As for the second integral in Eq.~\eqref{intsum}, assume the density of the classical FPT $\tau>0$ has strictly positive support (\emph{e.g.} Brownian search). In this case,
\begin{align}
    R_r(t) = \int_t^\infty e^{-rs}\mathbb{P}(\tau=s)\,\textup{d}s
    \leq ||\mathbb{P}(\tau=s)||_\infty \int_t^\infty e^{-rs}\,\textup{d}s
    = \frac{||\mathbb{P}(\tau=s)||_\infty}{r} e^{-rt}
\end{align}
which is finite for all $r>0$. Using this inequality, we have
\begin{align}
    \int_0^\infty tR_r(t)G_r(t)\,\textup{d}t \leq \frac{||\mathbb{P}(\tau=s)||_\infty}{r} \int_0^\infty t e^{-rt}G_r(t)\,\textup{d}t
    \leq \frac{||\mathbb{P}(\tau=s)||_\infty}{r} \int_0^\infty t e^{-rt}\,\textup{d}t
    = \frac{||\mathbb{P}(\tau=s)||_\infty}{r^3}.
\end{align}
The upper bound on Eq.~\eqref{intsum} in its entirety is then described by
\begin{align}
    \int_0^\infty tS_{\alpha r,r}(t)\,\textup{d}t \leq \frac{1-p_r}{\alpha rp_r}\Big( \frac{1}{\alpha r} + \frac{1-p_r}{\alpha rp_r} - \frac{\partial}{\partial r} \log p_r \Big) + \frac{||\mathbb{P}(\tau=s)||_\infty}{r^3p_r}.
\end{align}
Hence,
\begin{align} \label{frupper}
    f(r) \leq 2\Big( \frac{\alpha rp_r}{1-p_r} \Big)\Big( \frac{1}{\alpha r} + \frac{1-p_r}{\alpha rp_r} - \frac{\partial}{\partial r} \log p_r \Big) + \frac{2\alpha^2 p_r||\mathbb{P}(\tau=s)||_\infty}{r(1-p_r)^2}.
\end{align}
Taking $r\to\infty$ yields $\limsup_{r\to\infty} f(r)=2$. Combining the limits infimum and supremum gives the desired result. In Fig.~\ref{fig:cv} we numerically compute the CV by way of Eq.~\eqref{cv} and illustrate how it evolves with respect to the turnover rate $r$. Regardless of the scaling factor $\alpha$, we see that the CV converges to one in the high turnover limit.

\section{Pathwise interpretation of the universal bounds}

The universal bounds on the DRM MFPT have a simple pathwise interpretation. The key idea is to label searchers according to whether a recruited searcher, in the absence of other searchers, could eventually find the target before dying. This
labeling separates the time spent waiting for a potentially successful searcher from the time that this searcher subsequently needs to find the target.

Let $\sigma_n$ denote the recruitment time of searcher $n$, with $\sigma_0=0$ for the initial searcher and
\begin{equation}
\sigma_n=\xi_1+\cdots+\xi_n
\end{equation}
for $n\ge 1$ where $\xi_i\sim \textrm{Exp}(\lambda)$.
Let $\tau_n$ denote the intrinsic FPT of searcher $n$, measured from its own
recruitment time and let $D_n\sim \textrm{Exp}(\mu)$ denote its intrinsic lifetime. We assume the variables $\{\xi_i\}_{i\geq 0}$, $\{\tau_i\}_{i\geq 0}$, and $\{D_i\}_{i\geq 0}$ are mutually independent.

We call searcher $n$ successful if it finds the target before dying, \emph{i.e.}, if $\tau_n<D_n$. Now define
\begin{equation}
N:=\min\{n\ge 0:\tau_n<D_n\},
\label{eq:first_successful_index}
\end{equation}
the index of the first successful searcher in order of recruitment. Importantly, this is not necessarily the searcher that realizes the DRM FPT. A later recruit may be born after $\sigma_N$ and still overtake searcher $N$ by hitting the target sooner in absolute time. Nevertheless searcher $N$ gives a useful pathwise comparison.

Since no unsuccessful searcher finds the target before dying, and since no searcher can find
the target before it has been recruited, we have the pathwise bounds
\begin{equation}
\sigma_N \le T_{\lambda,\mu} \le \sigma_N+\tau_N,
\label{eq:pathwise_sandwich}
\end{equation}
which contain the probabilistic content of both bounds on the DRM MFPT.
\\
\subsection{The lower bound as the birth time of the first successful searcher}

The left inequality in Eq.~\eqref{eq:pathwise_sandwich} says that the DRM process cannot find
the target before the first eventually successful searcher has been recruited. Thus the lower
bound corresponds to an optimistic auxiliary process in which every searcher satisfying
$\tau_n<D_n$ is declared to find the target immediately upon recruitment. The success indicators of all the recruits are i.i.d.\ Bernoulli random variables
with parameter $p_\mu$ given by
\begin{equation}
p_\mu
=
\mathbb P(\tau<D)
=
\int_0^\infty e^{-\mu t} \mathbb P(\tau = t
)\,\textup{d}t.
\label{eq:prob_success_probability}
\end{equation}

Hence $N$ as in Eq.~\eqref{eq:first_successful_index} is geometric on $\{0,1,2,\dots\}$ with mean
\begin{align}
\mathbb E[N]
&=
\frac{1-p_\mu}{p_\mu}.
\end{align}
Since $N$ is independent of the recruitment times, we obtain
\begin{align}
\mathbb E[\sigma_N]
=\frac{1}{\lambda}\mathbb E[N]
=
\frac{1-p_\mu}{\lambda p_\mu}.
\end{align}
Taking expectations in the left inequality of Eq.~\eqref{eq:pathwise_sandwich} therefore gives
\begin{equation}
\mathbb E[T_{\lambda,\mu}]
\ge
\frac{1-p_\mu}{\lambda p_\mu}.
\label{eq:probabilistic_lower_bound}
\end{equation}

\subsection{The upper bound is obtained by following the first-born successful searcher}

The right inequality in Eq.~\eqref{eq:pathwise_sandwich} gives the upper bound. Once the
first successful searcher in recruitment order has been identified, one possible way for the
DRM process to find the target is simply to wait until this particular searcher finds it.
The actual process can only be faster, since any other successful searcher may also find the
target first.

Taking expectations in the right inequality of Eq.~\eqref{eq:pathwise_sandwich} gives
\begin{equation}
\mathbb E[T_{\lambda,\mu}]
\le
\mathbb E[\sigma_N]+\mathbb E[\tau_N].
\end{equation}
The first term is known. The second term is the mean hitting time
of a searcher conditioned on the event that it succeeds before dying. Indeed,
\begin{equation}
\mathbb P(\tau = t \mid \tau<D)
= \frac{e^{-\mu t}f_\tau(t)}{p_\mu}\,\textup{d}t.
\end{equation}
Hence
\begin{align}
\mathbb E[\tau_N]
=
\frac{\int_0^\infty t e^{-\mu t}f_\tau(t)\,\textup{d}t}{p_\mu}
=
-\frac{\partial_\mu p_\mu}{p_\mu}
=
-\partial_\mu \log p_\mu.
\end{align}
Combining this with the expression for $\mathbb E[\sigma_N]$ gives
\begin{equation}
\mathbb E[T_{\lambda,\mu}] \leq \frac{1-p_\mu}{\lambda p_\mu} - \partial_\mu \log p_\mu.
\label{eq:probabilistic_upper_bound}
\end{equation}

\section{Survival probability approach for the universal bounds}

\subsection{Optimal lower bound for the DRM MFPT}
The DRM MFPT can be expressed in terms of the corresponding survival probability,
\begin{equation} \label{firstmean}
    \mathbb{E}[T_{\lambda,\mu}]
    =
    \int_0^\infty S_{\lambda,\mu}(t)\,\mathrm{d}t.
\end{equation}
Since $S_{0,\mu}(t) \geq 1 - p_\mu$ for all $t>0$ with $p_\mu$ as defined in Eq.~\eqref{eq:prob_success_probability}, substitution into Eq.~\eqref{final} and subsequently into Eq.~\eqref{firstmean} reproduces the universal lower bound from Sect.~S2.A. given by
\begin{align}
    \mathbb{E}[T_{\lambda,\mu}] \geq \frac{1-p_\mu}{\lambda p_\mu}.
\end{align}

\subsection{Optimal upper bound for the DRM MFPT}
To determine an optimal upper bound on the MFPT, we first define a useful function, namely
\begin{equation} \label{Rmu}
    R_\mu(t)
    :=
    \int_t^\infty e^{-\mu t'}\, \mathbb P(\tau=t')\,\mathrm{d}t' ,
\end{equation} 
which is clearly bounded, $R_\mu(t)\in[0,p_\mu]$. It immediately follows that
\begin{equation} \label{s0mu}
    S_{0,\mu}(t)
    =
    1-p_\mu+R_\mu(t).
\end{equation}
Using Eq.~\eqref{s0mu} and defining $G_\mu(t)$ as the FPT survival probability of a DRM process initialized empty,
\begin{equation}
    G_\mu(t):=
    \exp\!\left(-\lambda\int_0^t \bigl(p_\mu-R_\mu(t')\bigr)\,\textrm{d}t'\right),
    \label{eq:Gdef}
\end{equation}
we can rewrite Eq.~\eqref{firstmean} as
\begin{align}
\mathbb{E}[T_{\lambda,\mu}]
&=
\frac{1-p_\mu}{p_\mu}
\int_0^\infty \bigl(p_\mu-R_\mu(t)\bigr)G_\mu(t)\,\mathrm{d}t
+
\frac{1}{p_\mu}\int_0^\infty R_\mu(t)G_\mu(t)\,\mathrm{d}t.
\label{eq:MFPT_decomp}
\end{align}
The first term in Eq.~\eqref{eq:MFPT_decomp} can be simplified,
\begin{align}
\frac{1-p_\mu}{p_\mu}
\int_0^\infty \bigl(p_\mu-R_\mu(t)\bigr)G_\mu(t)\,\mathrm{d}t
=
-\frac{1-p_\mu}{\lambda p_\mu}
\int_0^\infty G_\mu'(t)\,\mathrm{d}t =
\frac{1-p_\mu}{\lambda p_\mu}
\bigl(G_\mu(0)-G_\mu(\infty)\bigr) =
\frac{1-p_\mu}{\lambda p_\mu}.
\label{eq:first_term_exact}
\end{align}
For the second term, we use that $G_\mu(t)\in[0,1]$,
\begin{equation}
\frac{1}{p_\mu}\int_0^\infty R_\mu(t)G_\mu(t)\,\mathrm{d}t
\le
\frac{1}{p_\mu}\int_0^\infty R_\mu(t)\,\mathrm{d}t,
\end{equation}
and finally that
\begin{align}
    \int_0^\infty R_\mu(t)\,\mathrm{d}t
    =
    -\frac{\partial p_\mu}{\partial \mu}.
    \label{eq:q_derivative}
\end{align}
Combining the preceding equations gives
\begin{equation}
    \mathbb{E}[T_{\lambda,\mu}]
    \le
    \frac{1-p_\mu}{\lambda p_\mu}
    -
    \frac{\partial}{\partial \mu}\log p_\mu,
    \label{eq:upper_bound}
\end{equation}
reproducing the universal upper bound from Sect.~S2.B.

\section{When does $-\partial_\mu \log p_\mu \to 0$?}

Based on the probabilistic derivation of the bounds, the second term in the upper bound,
\begin{equation}
    -\frac{\partial}{\partial \mu}\log p_\mu,
\end{equation}
has a simple interpretation. Indeed, it is the mean first-passage time of a single searcher conditioned on finding the target before dying. This interpretation also allows us to see when the second term vanishes. Let
\begin{equation}
    t_* :=
    \inf\{t\ge 0: \mathbb P(\tau = t)>0\}
\end{equation}
be the left edge of the support of the first-passage time density. As $\mu\to\infty$, the exponential tilt of the conditioned first-passage time density increasingly selects the fastest possible successful trajectories, and hence
\begin{equation}
    -\frac{\partial}{\partial \mu}\log p_\mu
    \to 
    t_*.
    \label{eq:second_term_limit}
\end{equation}
Therefore, whenever there is no hard cutoff on travel time to the target and arbitrarily short first-passage times are possible (e.g. in Brownian diffusion), $t_* \to 0$, and consequently, 
\begin{equation}
    -\frac{\partial}{\partial \mu}\log p_\mu \to 0.
\end{equation}
For example, for Brownian motion on the half-line, where $p_\mu=e^{-x_0\sqrt{\mu/D}}$, we have
\begin{equation}
    -\frac{\partial}{\partial \mu}\log p_\mu = \frac{x_0}{2\sqrt{D\mu}} \to 0.
\end{equation}
By contrast, if the intrinsic dynamics impose a strictly positive minimum arrival time $t_*>0$, then the second term does not vanish. Instead,
\begin{equation}
    -\frac{\partial}{\partial \mu}\log p_\mu
    \to
    t_*.
\end{equation}
The simplest example is deterministic first passage: if $\tau\equiv t_*$ (or $\mathbb P(\tau = t) = \delta(t-t_*)$), then $p_\mu=e^{-\mu t_*}$, and hence
\begin{equation}
    -\frac{\partial}{\partial \mu}\log p_\mu=t_*.
\end{equation}

\section{When DRM outpaces stochastic resetting} \label{outpace}
Consider again that $\lambda = \alpha r$ and $\mu=r$. The purpose of this section is to provide a detailed derivation of two results discussed in the main text. 
\begin{enumerate}
    \item In the high-turnover limit, the ratio of the DRM MFPT to the resetting MFPT satisfies,
    \begin{equation}
        \lim_{r\to\infty}\frac{\mathbb E[T_{\alpha r,r}]}{\mathbb E[\tau_r]} = \frac{1}{\alpha}.
    \end{equation}
    This establishes quite generally that stochastic resetting stops being a lower bound to DRM for any $\alpha>1$. Moreover, there exists a critical value $r^*$, such that for all $r>r^*$, we have $\mathbb E[T_{\alpha r,r}] < \mathbb E[\tau_r]$.
    \item For the special case of 1D Brownian motion, when the DRM search is sufficiently redundancy dominated, the DRM MFPT becomes a lower bound to the stochastic resetting MFPT for all $r$. More precisely, we have 
    \begin{equation}
       \mathbb E[\tau_r]  \geq \mathbb E[T_{\alpha r,r}] \quad \textup{for all } \alpha > \phi
    \end{equation}
    where $\phi = \frac{1+\sqrt{5}}{2} \approx 1.618$ is the golden ratio -- a quantity that emerges in diverse physical and mathematical contexts.
\end{enumerate}

\subsection{When resetting stops being a lower bound to DRM}
Recall that the DRM MFPT can be described by
\begin{equation}
\mathbb E[T_{\lambda,\mu}]
=
\int_0^\infty S_{\lambda,\mu}(t)\,\textup{d}t
\end{equation}
with $S_{\lambda,\mu}(t)$ as in Eq.~\eqref{final}. For convenience, we define
\begin{equation}
q_r
=
\int_{0}^{\infty} R_r(t)\,\textup{d}t
\end{equation}
with $R_r(t)$ as in Eq.~\eqref{Rmu}. Substitution of $S_{0,r}(t)$ as in Eq.~\eqref{s0mu} into Eq.~\eqref{final} with $\lambda=\alpha r$ yields
\begin{equation}
(1-p_r)e^{-\alpha r p_r t}
\le
S_{\alpha r,r}(t)
\le
e^{\alpha r q_r}(1-p_r+R_r(t))e^{-\alpha r p_r t}.    
\end{equation}
Integrating over time yields
\begin{equation} 
\label{Earr}
\frac{1-p_r}{\alpha r p_r}
\le
\mathbb E[T_{\alpha r,r}]
\le
e^{\alpha r q_r}
\left(
\frac{1-p_r}{\alpha r p_r}+q_r
\right),
\end{equation}
where we used $e^{-\alpha r p_r t} \le 1$ for all $t\ge0$. It remains to show that $q_r$ has only a subleading contribution to Eq.~\eqref{Earr} as $r\to\infty$. Through integration by parts, we have
\begin{equation}
q_r=\int_0^\infty t\,e^{-rt}\mathbb{P}(\tau=t)\,\textup{d}t,
\end{equation}
and thus
\begin{equation}
r q_r=\int_0^\infty rt e^{-rt}\mathbb{P}(\tau=t)\,\textup{d}t.
\end{equation}
For any $\varepsilon>0$, there exists $\delta>0$ such that
\begin{equation}
\int_0^\delta \mathbb P(\tau=t)\,\text{d}t<\varepsilon.
\end{equation}
Thus, for $r>\delta^{-1}$, $rte^{-rt}$ is decreasing on
$[\delta,\infty)$, and hence
\begin{align}
0\le rq_r \le
e^{-1}\int_0^\delta \mathbb P(\tau=t)\,\textup{d}t +
r\delta e^{-r\delta}\int_\delta^\infty \mathbb P(\tau=t)\,\textup{d}t \le e^{-1}\varepsilon+r\delta e^{-r\delta}.
\end{align}
Taking $r\to\infty$ and then $\varepsilon\to0$ gives $rq_r\to0$. The same cutoff argument gives $p_r\to0$. Since
$\mathbb E[\tau_r]=(1-p_r)/(rp_r)$, Eq.~\eqref{Earr} implies
\begin{equation}
\frac{1}{\alpha} \leq \frac{\mathbb E[T_{\alpha r,r}]}{\mathbb E[\tau_r]} \leq e^{\alpha rq_r}\left(\frac{1}{\alpha}+\frac{rp_rq_r}{1-p_r}\right).
\end{equation}
As $rq_r\to0$ and $p_r\to0$, the right-hand side converges to
$1/\alpha$. Therefore,
\begin{equation}
\lim_{r\to\infty}\frac{\mathbb E[T_{\alpha r,r}]}{\mathbb E[\tau_r]} = \frac{1}{\alpha}.  
\end{equation}

\subsection{When DRM becomes a lower bound to stochastic resetting}

Consider Brownian motion on the half-line with diffusion coefficient $D>0$, initial position $x_0>0$, and a target at the origin. Set
\begin{equation} \label{zv}
    z:=x_0\sqrt{\frac{r}{D}},
    \qquad
    v:=rt.
\end{equation}
The survival probability of a mortal Brownian searcher with mortality rate $r>0$ is given by
\begin{equation}
S_{0,r}\!\left(\frac{v}{r}\right)
=
1-\frac{e^{-z}}{2}
\left[
2-
\operatorname{erfc}\!\left(\sqrt v-\frac{z}{2\sqrt v}\right)
+
e^{2z}\operatorname{erfc}\!\left(\sqrt v+\frac{z}{2\sqrt v}\right)
\right]
\label{eq:brownian_mortal_survival}
\end{equation}
where $\textup{erfc}(x)$ denotes the complementary error function. For fixed $v>0$,  $S_{0,r}$ has the small-$z$ expansion
\begin{equation}
    S_{0,r}\!\left(\frac{v}{r}\right)
    =
    z\left[
    \operatorname{erf}(\sqrt v)+\frac{e^{-v}}{\sqrt{\pi v}}
    \right]
    +O(z^2)
    \label{eq:S0_small_z}
\end{equation}
with $\textup{erf}(x)$ denoting the (standard) error function. Taking $\lambda=\alpha r$ and $\mu=r$,
\begin{equation}
S_{\alpha r,r}\!\left(\frac{v}{r}\right)
=
S_{0,r}\!\left(\frac{v}{r}\right)
\exp\!\left[-\alpha\int_0^v
\left(1-S_{0,r}\!\left(\frac{t'}{r}\right)\right)\,\mathrm{d}t'\right].
\end{equation}
Using Eq.~\eqref{eq:S0_small_z}, the exponent is $-\alpha v+O(z)$ for fixed $v$, and therefore
\begin{equation}
S_{\alpha r,r}\!\left(\frac{v}{r}\right)
=
ze^{-\alpha v}
\left[
    \operatorname{erf}(\sqrt v)+\frac{e^{-v}}{\sqrt{\pi v}}
\right]
+O(z^2).
\end{equation}
Integrating over time yields
\begin{align} \label{eq:brownian_Ialpha}
    \mathbb{E}[T_{\alpha r,r}]
    =
    \frac{z}{r} \int_0^\infty
    \left[
    \operatorname{erf}(\sqrt v)+\frac{e^{-v}}{\sqrt{\pi v}}
    \right]
    e^{-\alpha v}\,\mathrm{d}v +o\!\left(\frac{z}{r}\right) =
    \frac{z}{r} \frac{\sqrt{\alpha+1}}{\alpha}+o\!\left(\frac{z}{r}\right)
\end{align}
and hence by substitution of $z$ in Eq.~\eqref{zv} we have
\begin{equation}
    \lim_{r\to 0} \frac{\sqrt{Dr}}{x_0}\mathbb{E}[T_{\alpha r,r}] = \frac{\sqrt{\alpha+1}}{\alpha}.
    \label{eq:brownian_low_r_DRM}
\end{equation}
Since $p_r = \textup{exp}(-x_0\sqrt{r/D})$ for Brownian motion with rate-$r$ stochastic resetting, it follows that
\begin{equation}
    \mathbb{E}[\tau_r]=\frac{1-p_r}{r p_r} \sim
    \frac{x_0}{\sqrt{Dr}} \label{eq:brownian_low_r_resetting}
\end{equation}
as $r\to 0$ and thus 
\begin{equation}
    \lim_{r\to 0} \frac{\mathbb{E}[T_{\alpha r,r}]}{\mathbb{E}[\tau_r]}
    =
    \frac{\sqrt{\alpha+1}}{\alpha}.
    \label{eq:low_r_ratio}
\end{equation}
The leading low-turnover comparison changes sign when the right-hand side of Eq.~\eqref{eq:low_r_ratio} equals one, that is, when
\begin{equation}
    \alpha = \phi =\frac{1+\sqrt{5}}{2}.
    \label{eq:golden_ratio}
\end{equation}
The golden ratio $\phi$ is therefore the low-turnover endpoint of the Brownian phase boundary separating the region where the DRM MFPT outpaces that of stochastic resetting from that in which the resetting MFPT outpaces DRM.

\end{document}